# A Survey of Energy-Efficient Techniques for 5G Networks and Challenges Ahead

Stefano Buzzi, *Senior Member, IEEE,* Chih-Lin I, *Senior Member, IEEE,* Thierry E. Klein, *Member, IEEE,* H. Vincent Poor, *Fellow, IEEE,* Chenyang Yang, *Senior Member, IEEE,* and Alessio Zappone, *Member, IEEE*

*Abstract*—After about a decade of intense research, spurred by both economic and operational considerations, and by environmental concerns, energy efficiency has now become a key pillar in the design of communication networks. With the advent of the fifth generation of wireless networks, with millions more base stations and billions of connected devices, the need for energy-efficient system design and operation will be even more compelling. This survey provides an overview of energy-efficient wireless communications, reviews seminal and recent contribution to the state-of-the-art, including the papers published in this special issue, and discusses the most relevant research challenges to be addressed in the future.

*Index Terms*—Energy efficiency, 5G, resource allocation, dense networks, massive MIMO, small cells, mmWaves, visible-light communications, cloud RAN, energy harvesting, wireless power transfer.

## I. INTRODUCTION

ENERGY consumption has become a primary concern in the design and operation of wireless communication systems. Indeed, while for more than a century communication networks have been mainly designed with the aim of optimizing performance metrics such as the data-rate, throughput, latency, etc., in the last decade energy efficiency has emerged as a new prominent figure of merit, due to economic, operational, and environmental concerns. The design of the next generation (5G) of wireless networks will thus necessarily have to consider energy efficiency as one of its key pillars. Indeed, 5G systems will serve an unprecedented number of devices, providing ubiquitous connectivity as well as innovative and rate-demanding services. It is forecast that by 2020 there will be more than 50 billion connected devices [1], i.e. more that 6 connected devices per person, including not only human-type communications, but also machine-type communications. The vision is to have a connected society in which sensors,

S. Buzzi is with the Department of Electrical and Information Engineering at the Università di Cassino e del Lazio Meridionale, Cassino, Italy (buzzi@unicas.it).

C.-L. I is with the Green Communication Research Center, China Mobile Research Institute, Beijing 100053, China (e-mail: icl@chinamobile.com).

T. E. Klein is with Nokia Innovation Steering, Nokia, Murray Hill, NJ 07974 USA (thierry.klein@nokia.com).

C. Yang is with the School of Electronics and Information Engineering at Beihang University, Beijing, China (cyyang@buaa.edu.cn).

H. V. Poor is with the Department of Electrical Engineering, Princeton University, Princeton, NJ 08544 USA (poor@princeton.edu).

A. Zappone is with the Communication Department of the Technische Universität Dresden, Dresden, Germany (Alessio.Zappone@tu-dresden.de).

This research was supported in part by the U. S. National Science Foundation under Grants ECCS-1343210 and ECCS-1549881, and also by the German Research Foundation (DFG) project CEMRIN, under grant ZA 747/1-3.

cars, drones, medical and wearable devices will all use cellular networks to connect with one another, interacting with human end-users to provide a series of innovative services such as smart homes, smart cities, smart cars, telesurgery, and advanced security. Clearly, in order to serve such a massive number of terminals, future networks will have to dramatically increase the provided capacity compared to present standards. It is estimated that the traffic volume in 5G networks will reach tens of Exabytes ($1000^6$ Bytes) per month. This requires the capacity provided by 5G networks to be 1000 times higher than in present cellular systems [2]. Trying to achieve this ambitious goal relying on the paradigms and architectures of present networks is not sustainable, since it will inevitably lead to an energy crunch with serious economic and environmental concerns.

**Economic concerns.** Current networks are designed to maximize the capacity by scaling up the transmit powers. However, given the dramatic growth of the number of connected devices, such an approach is not sustainable. Using more and more energy to increase the communication capacity will result in unacceptable operating costs. Present wireless communication techniques are thus simply not able to provide the desired capacity increase by merely scaling up the transmit powers.

**Environmental concerns.** Current wireless communication systems are mainly powered by traditional carbon-based energy sources. At present, information and communication technology (ICT) systems are responsible for 5% of the world's $CO_2$ emissions [3], [4], but this percentage is increasing as rapidly as the number of connected devices. Moreover, it is foreseen that 75% of the ICT sector will be wireless by 2020 [5], thus implying that wireless communications will become the critical sector to address as far as reducing ICT-related $CO_2$ emissions is concerned.

### A. Averting the energy crunch

In order to avert the energy crunch, new approaches to wireless network design and operation are needed. The key point on which there is general consensus in the wireless academic and industry communities, is that the $1000\times$ capacity increase must be achieved *at a similar or lower power consumption as today's networks* [6], [7]. This means that the efficiency with which each Joule of energy is used to transmit information must increase by a factor 1000 or more. Increasing the network energy efficiency has been the goal of the GreenTouch consortium [8], which was founded in 2010 as an open global pre-competitive research consortium with the focus to improve network energy efficiency by a factor 1000 with respect to the 2010 state of the art reference network. The consortium published a technology roadmap and announced its final results in its "Green Meter" research study [9].

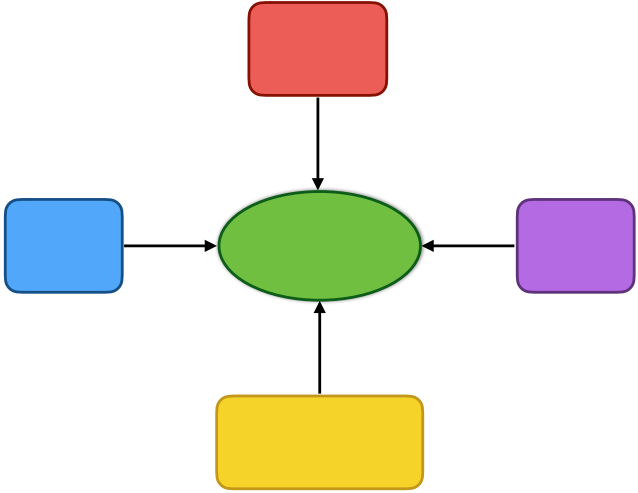

Figure 1. Energy-efficient 5G technologies.

Additionally, the Groupe Speciale Mobile Association (GSMA) demands, by 2020, a reduction of $CO_2$ emissions per connection of more than 40%. These fundamental facts have led to introducing the notion of bit-per-Joule energy efficiency, which is defined as the amount of information that can be reliably transmitted per Joule of consumed energy, and which is a key performance indicator for 5G networks [6], [7] (see also [10]–[12] as some of the first papers introducing the notion of bit-per-Joule energy efficiency). As illustrated in Fig. 1, most of the approach useful for increasing the energy efficiency of wireless networks can be grouped under four broad categories as follows.

**a) Resource allocation.** The first technique to increase the energy efficiency of a wireless communication system is to allocate the system radio resources in order to maximize the energy efficiency rather than the throughput. This approach has been shown to provide substantial energy efficiency gains at the price of a moderate throughput reduction [13].

**b) Network planning and deployment.** The second technique is to deploy infrastructure nodes in order to maximize the covered area per consumed energy, rather than just the covered area. In addition, the use of base station (BS) switch-on/switch-off algorithms and antenna muting techniques to adapt to the traffic conditions, can further reduce energy consumptions [14], [15].

**c) Energy harvesting and transfer.** The third technique is to operate communication systems by harvesting energy from the environment. This applies to both renewable and clean energy sources like sun or wind energy, and to the radio signals present over the air.

**d) Hardware solutions** The fourth technique is to design the hardware for wireless communications systems explicitly accounting for its energy consumption [16], and to adopt major architectural changes, such as the cloud-based implementation of the radio access network [17].

In the following, a survey of the state-of-the-art relative to the above cited four categories is given, with a special focus on the papers published in this issue.

## II. RESOURCE ALLOCATION

As energy efficiency has emerged as a key performance indicator for future 5G networks, a paradigm shift from throughput-optimized to energy-efficiency-optimized communications has begun. A communication system's radio resources should no longer be solely optimized to maximize the amount of information that is reliably transmitted, but rather the amount of information that is reliably transmitted per Joule of consumed energy. Compared to traditional resource allocation schemes, this requires the use of novel mathematical tools specifically tailored to energy efficiency maximization. A survey of this topic is provided in [13].

From a physical standpoint, the efficiency with which a system uses a given resource, is the ratio between the benefit obtained by using the resource, and the corresponding incurred cost. Applying this general definition to communication over a wireless link, the cost is represented by the amount of consumed energy, which includes the radiated energy, the energy loss due to the use of non-ideal power amplifiers, as well as the static energy dissipated in all other hardware blocks of the system (e.g. signal up and downconversion, frequency synthesizer, filtering operations, digital-to-analog and analog-to-digital conversion, and cooling operations). In the literature it is usually assumed that the transmit amplifiers operate in the linear region, and that the static hardware energy is independent of the radiated energy [4], [13], [18]–[20]. These two assumptions lead to expressing the consumed energy during a time interval $T$ as

$$E = T(\mu p + P_c) \text{ [Joule]} , \quad (1)$$

wherein $p$ is the radiated power, $\mu = 1/\eta$, with $\eta$ the efficiency of the transmit power amplifier, and $P_c$ includes the static power dissipated in all other circuit blocks of the transmitter and receiver.

On the other hand, the benefit produced after sustaining the energy cost in (1) is related to the amount of data reliably transmitted in the time interval $T$, and several performance functions have been employed in the literature to measure this quantity, depending on the particular system under analysis. Some notable examples are:

- **System capacity / Achievable rate.** The Shannon capacity (or the achievable rate in scenarios where the capacity is not known), often expressed through the well-known formula involving the log of one plus the Signal-to-Interference plus Noise-Ratio (SINR), represents the ultimate rate at which reliable communication is possible, and therefore captures the needs of having both fast and reliable communication. This measure has been considered recently in [18] and [21], which focused mainly on multi-carrier systems. Subsequent to these works, the achievable rate has become the dominant choice to measure the quality of a communication system. Further studies that considered this approach are [22], and [23] for OFDMA systems, [24]–[26] for MIMO systems, [27]–[29] for relay-assisted communications, [30] and [31] for cognitive communications.

- **Throughput.** The system throughput is a measure that, differently from capacity, takes into account the actual rate at which data is transmitted. Its consideration however requires specifying the system bit error rate, and thus the particular modulation in use. The throughput was the first metric to be considered in the context of energy efficiency, and its use dates back to the seminal works [11], [12], and [32], in the context of CDMA networks. These studies spurred the interest for energy-efficient resource allocation in wireless networks, and a throughput-based definition of energy efficiency was used in [33]–[36] proposing game-theoretic approaches for energy-efficient multiple access networks, in [37] with reference to cognitive radio systems, in [38] in conjunction with widely linear receivers, in [39] for ultra-wideband systems, and in [40] for relay-based systems.

- **Outage capacity.** The above metrics refer to scenarios in which perfect channel state information (CSI) is available at the resource allocator. Also, they can be replaced by their ergodic counterparts in fast-fading scenarios, or in general whenever only statistical CSI is available at the resource allocator [41]. Instead, in slow-fading scenarios, outage events become the major impairments of the communication and the outage capacity becomes the most suitable metric to measure the benefit obtained from the system. This approach is somewhat less popular than the ones discussed above. Nevertheless, it has attracted significant interest and recent contributions in this area can be found in [42] and [43].

A common feature of all the above metrics is that they are measured in [bits/s] and depend on the signal-to-noise ratio (SNR) (or SINR) of the communication, denoted by $\gamma$. Thus, we can generally express the system benefit by a function $f(\gamma)$, with $f$ to be specified according to the particular system to optimize.

Finally, we can define the energy efficiency of a communication link as

$$EE = \frac{Tf(\gamma)}{T(\mu p + P_c)} = \frac{f(\gamma)}{\mu p + P_c} \text{ [bits/Joule] .} \qquad (2)$$

It can be seen that (2) is measured in [bits/Joule], thus naturally representing the efficiency with which each Joule of energy is used to transmit information. Fig. 2 shows the typical shape of the energy efficiency versus the transmit power, for different values of the static power consumption $P_c$.

Two important observations can be made from Fig. 2.

1) First, the energy efficiency is not monotone in the transmit power and is maximized by a finite power level. This is a fundamental difference compared to traditional performance metrics, which instead are monotonically increasing in the transmit power. While the maximization of the numerator of the energy efficiency leads to transmitting with a power level equal to the maximum feasible

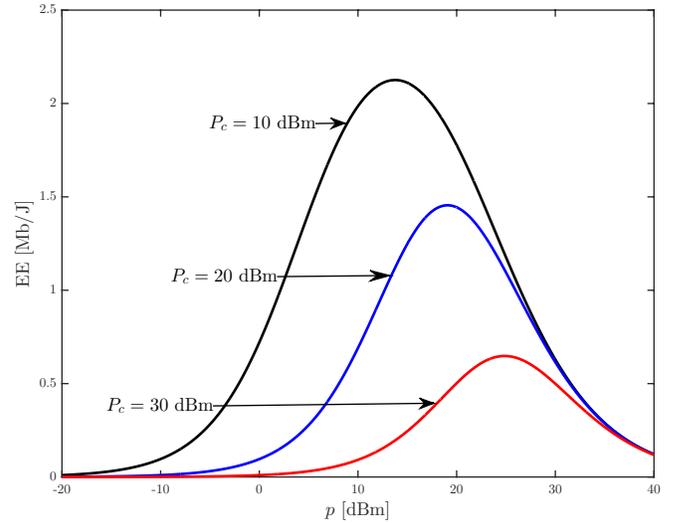

Figure 2. Typical shape of the energy efficiency function.

transmit power, maximizing the energy efficiency yields a power level that is in general lower.

2) Increasing the static power term causes the maximizer of the energy efficiency to increase. In the limit $P_c >> p$, the denominator becomes approximately a constant, and energy efficiency maximization reduces to the maximization of the numerator.

The energy efficiency in (2) refers to a single communication link. In a communication network the expression becomes more involved, depending on the benefits and costs incurred by each individual link of the network and several network-wide performance functions have been proposed. Two main approaches can be identified.

- **Network benefit-cost ratio.** The network benefit-cost ratio is given by the ratio between the sum of all individual benefits of the different links, and the total power consumed in the network. This metric is called *Global Energy Efficiency* (GEE) and is the network-wide energy efficiency measure with the strongest physical meaning. A considerable number of contributions have provided schemes for GEE maximization. Among recent examples we mention [22], [27], [44] and [23] for OFDMA networks, and [26], [45], [46] and [47] for MIMO systems. Among these references, [27], [46] and [47] also consider the presence of relays.

- **Multi-objective approach.** One drawback of the GEE is that it does not allow tuning of the individual energy efficiencies of the different network nodes. To address this issue, an alternative approach is to regard each individual node's energy efficiency as a different objective to maximize, thus performing a multi-objective resource allocation, maximizing a combination of the different energy efficiencies $\{EE_k\}_{k=1}^K$ according to some increasing function $\phi(EE_1, \ldots, EE_K)$. Several combining functions $\phi$ have been proposed:

1) *Weighted Sum Energy Efficiency* (WSEE). The function $\phi$ is the weighted sum of the individual energy efficien-

cies. Contributions using this approach are [23], [48] and [49].

2) *Weighted Product Energy Efficiency* (WPEE). The function $\phi$ is the weighted product of the individual energy efficiencies. This approach is related to the proportional fairness criterion, and studies that have considered this metric are [23] and [50].

3) *Weighted Minimum Energy Efficiency* (WMEE). The function $\phi$ is the weighted minimum of the individual energy efficiencies. This approach corresponds to a worst-case design, and has been considered in [47] and [51].

All three combining functions are able to describe (at least) parts of the energy-efficient Pareto boundary of the system, by varying the choice of the weights. However, only the WMEE is able to describe the complete Pareto boundary, by sweeping the weights [13].

Energy efficiency maximization can be also carried out subject to all practical constraints that are typically enforced in communication systems. Besides the widely considered maximum power constraint, more recently quality of service (QoS) constraints have started being enforced. This includes minimum rate guarantees [47], [52], minimum delay constraints [53], [54], maximal delay bound constraints [55], and interference temperature constraints, typically enforced in underlay systems [30], [56].

It should also be mentioned that an alternative, yet less powerful approach for energy saving is, rather than maximizing the energy efficiency, to minimize the energy consumption (see [57] and [58] for recent contributions in this direction). Although the two approaches might seem equivalent, they in general lead to different resource allocations. Of course, in order to rule out the trivial zero-power solution, the consumed energy minimization problem must be coupled with some minimum QoS constraints to be guaranteed in terms of capacity, outage capacity, or throughput. This approach results in consuming the minimum amount of energy required to maintain given minimum system performance; however, it does not account for the system benefit-cost ratio, and therefore might be overly pessimistic, being able only to provide minimum acceptable performance. Also, power minimization is a particular case of energy efficiency maximization subject to QoS constraints, which corresponds to minimizing the denominator of the energy efficiency, with a constraint on the numerator.

In this special issue, energy-efficient resource allocation is studied in [59]–[64]. In [59], weighted sum energy efficiency maximization is considered. A distributed algorithm is provided, by means of pricing and fractional optimization techniques. In [60] energy-efficient licensed-assisted access for LTE systems has been proposed to use unlicensed bands. The energy efficiency of the system is studied and the energy-efficient Pareto region is characterized. A Lyapunov optimization technique is employed in [61] to optimize the energy efficiency of WiFi networks with respect to network selection, sub-channel assignment, and transmit power. Energy efficiency optimization in MIMO-OFDM networks is performed in [62], considering the practical scenario in which the propagation channels vary dynamically over time (for example due to user mobility, fluctuations in the wireless medium, and changes in the users' loads). Learning tools are combined with fractional programming to develop an online optimization algorithm. The energy efficiency maximization of a massive MIMO system operating in the mmWave range is studied in [63], while the paper [64] focuses on a cognitive scenario wherein a secondary network co-exists, using the same frequency band, with a primary cellular network.

## III. NETWORK PLANNING AND DEPLOYMENT

In order to cope with the sheer number of connected devices, several potentially disruptive technologies have been proposed for the planning, deployment, and operation of 5G networks.

### A. Dense networks

The idea of dense networks is to deal with the explosively increasing number of devices to serve by increasing the amount of deployed infrastructure equipment. Two main kinds of network densification are gaining momentum and appear as very strong candidates for the implementation of 5G networks.

*1) Dense Heterogeneous Networks:* Unlike present network deployments which uniformly split a macro-cell into a relatively low number of smaller areas each covered by a light base-station, dense heterogeneous networks drastically increase the number of infrastructure nodes per unit of area [7], [65]. A very large number of heterogeneous infrastructure nodes ranging from macro BSs to femto-cells and relays are opportunistically deployed and activated in a demand-based fashion, thus leading to an irregularly-shaped network layout such as that shown in Fig. 3.

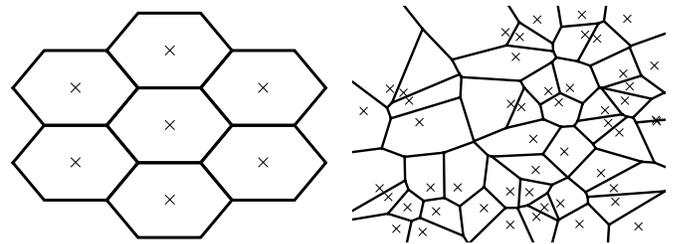

Figure 3. Evolution of cellular network layout. Traditional layout on the left; 5G layout on the right.

One critical challenge when dealing with dense heterogeneous networks is the modeling of the positions of the nodes in the network, which is typically difficult to predict deterministically. Instead, the nodes' locations are modeled as random variables with a given spatial distribution, and in this context the most widely used tool is the theory of *stochastic geometry* [66], [67]. Employing this tool, most research effort on dense networks has been focused on the analysis of traditional, non-energy-efficient performance measures [68], [69].

Fewer results are available as far as energy efficiency is concerned. From an energy-efficient point-of-view, node densification reduces the (electrical and/or physical) distances between communicating terminals, thus leading to higher data-rates at lower transmit powers. However, it also creates additional interference which might degrade the network energy efficiency. This trade-off has been analyzed in [70], where it is shown that densification has a beneficial impact on energy efficiency, but the gain saturates as the density of the infrastructure nodes increases, thus indicating that an optimal density level exists. The optimal network densification level is investigated also in [71], where a threshold value on the operating cost of a small BS is determined. If below the threshold, micro BSs are beneficial, otherwise they should be switched off. In [72], fractional programming is employed to develop a spectrum allocation algorithm in OFDMA heterogeneous networks, so as to minimize the energy expenditure per transmitted bit. The paper [73], in this special issue, shows, using a game-theoretic approach, that infrastructure sharing between different mobile network operators (MNOs) may bring substantial energy savings by increasing the percentage of BSs in sleep mode. In particular, the paper considers two different MNOs coexisting in the same area, which are aggregated as a single group to make the day-ahead and real-time energy purchase, while their BSs share the wireless traffic to maximally turn lightly-loaded BSs into sleep mode.

*2) Massive MIMO:* If the idea of dense networks is to densify the number of infrastructure nodes, the idea of massive MIMO is to densify the number of deployed antennas. In massive MIMO, conventional arrays with only a few antennas fed by bulky and expensive hardware are replaced by hundreds of small antennas fed by low-cost amplifiers and circuitry. The research interest in such a technology has been spurred by [74], which observed how, owing to the law of large numbers, large antenna arrays can average out multi-user interference. This happens provided the so-called *favorable propagation condition* holds, which has been experimentally validated in the overview works [75] and [76].

However, massive MIMO systems also come with several challenges and impairments. First of all, deploying a very large number of antennas points in the direction of very large systems, for which a microscopic analysis is usually too complex. Instead, system analysis and design must be performed based on the limiting behavior of the network, a task which is usually accomplished by means of *random matrix theory* [77], [78]. In addition, massive MIMO systems are characterized by a more difficult channel estimation task, due to a more severe pilot contamination effect, and to more significant hardware impairments. Contributions to address these challenges have mainly focused on traditional performance measures (see for example [79] and [80]), and results on the energy efficiency of massive MIMO systems have started appearing only very recently.

As far as energy efficiency is concerned, massive MIMO has been shown to reduce the radiated power by a factor proportional to the square root of the number of deployed antennas, while keeping the information rate unaltered [81]. However, this result applies to an ideal, single-cell massive MIMO system only, and without taking into account the hardware-consumed power. In [82], the aggregate effect of hardware impairments in massive MIMO systems is accounted for and the energy efficiency is analyzed. In [47] and [83] the hardware power is included in the analysis, and it is shown that the network energy efficiency is maximized for a finite number of deployed antennas.

Contributions [84]–[88] in this special issue address the topic of energy efficiency in dense networks. Contribution [84] presents a framework for self-organizing cells that autonomously activate or deactivate in response to traffic demands. Network deployment in response to traffic conditions is also analyzed in [85]. There, it is shown that in the presence of heavy traffic conditions, deploying indoor small cells is more energy-efficient than traditional network layouts. In [86] an optimization framework for energy-efficient radio resource management in heterogeneous networks is developed, assuming stochastic traffic arrivals. Reference [87] merges the heterogeneous approach and the massive MIMO technology, studying the problem of determining the optimal BS density, transmit power levels, and number of deployed antennas for maximal energy efficiency. Alternatively, [88] considers a two-way relay channel, in which multiple pairs of full-duplex users exchange information through a full-duplex amplify-and-forward relay with massive antennas. Different transceiver strategies are analyzed, which are shown to achieve energy efficiency gains.

### B. Offloading techniques

Offloading techniques are another key 5G strategy instrumental to boost the capacity and energy efficiency of future networks. Currently available user devices are already equipped with multiple radio access technologies (RATs) – e.g., cellular, Bluetooth, WiFi –, so that, whenever alternative connection technologies are available (e.g., as often happens in indoor scenarios), cellular traffic can be offloaded and additional cellular resources can be provided to those users that cannot offload their traffic. Future networks will vastly rely on offloading techniques, and these will not only be based on Wi-Fi. In particular, the following offloading methods/strategies can be envisioned:

- **Device-to-device (D2D) communications.** While in a conventional network user devices are not allowed to directly communicate, D2D communications [89], [90] refer instead to the scenario in which several co-located (or in close proximity) devices can communicate directly *using a cellular frequency and being instructed to do so by the BS*. D2D techniques have a profound impact on the system energy efficiency since direct transmission between nearby devices may happen at a much lower transmit power than that needed for communication through a BS that can be far away. Additionally, they are a powerful offloading strategy since they permit releasing resources at the BS that, through proper interference management, can be used for supporting other users. The impact of D2D communications on the energy efficiency of future wireless networks has been studied in [91]–[94].
- **Visible light communications (VLC).** VLC, also known as LiFi or optical wireless communication (OWC), is a technology that can serve indoor communications in future wireless systems. While being basically a short range technology, it has some remarkable advantages, such as very high energy-efficiency, availability of large bandwidths and thus the capability to support large data-rates [95]. The use of the visible light spectrum for data communication is enabled by inexpensive and off-the-shelf available light emitting diodes (LEDs). Individual LEDs can be modulated at very high speeds, and indeed 3.5Gbit/s@2m distance has been demonstrated as well as 1.1Gbit/s@10m, both with a total optical output power of 5 mW [96], [97].

- **Local caching.** Wireless networks are subject to time-varying traffic loads, and light load periods can be leveraged by using the redundant capacity to download and store in the BS's cache popular content that is likely to be requested by several users. This strategy, named "local caching", basically trades off storage capacity (a quite inexpensive resource) at the BS with network throughput. By reducing the load on the backhaul link, content caching strategies can potentially increase the energy efficiency of the core network by avoiding multiple transmissions of the same content for different users. Local caching is a relatively new offloading technique, which however has been attracting a growing interest [98]–[101].

- **mmWave cellular.** The use of frequency bands above 10GHz, a.k.a. mmWaves [102], while increasing the available network bandwidth, is considered in this section as a strategy to offload traffic from the sub-6GHz cellular frequencies for short-range (up to 100-200 m) communications in densely crowded areas. Future wireless technology will need to harness the massively unused mmWave spectrum to meet the projected acceleration in mobile traffic demand. Today, the available range of mmWave-based solutions is already represented by IEEE 802.11ad (WiGig), IEEE 802.15.3c, WirelessHD, and ECMA-387 standards, with more to come in the following years. In Section V we will comment on the hardware challenges that the use of mmWave poses. Studies on the impact of mmWaves on the energy efficiency of future 5G can be found in [103]–[105].

In this special issue, offloading approaches for energy efficiency are investigated in the papers [106]–[109] and [110].

The paper [106] targets energy-aware D2D communications underlaying a cellular system, and investigates several fundamental problems, including the potential energy savings of D2D communications, the underlying reasons for the savings and the tradeoff between energy consumption and other network parameters such as available bandwidth, buffer size and service delay in large scale D2D communication networks. By formulating a mixed integer linear programming problem that minimizes the energy consumption for data transmission from the cellular BSs to the end-user devices through any possible ways of transmissions, a theoretical performance lower bound of system energy consumption is obtained, showing the energy savings granted by D2D communications.

Papers [107] and [108] focus on VLC. The former manuscript investigates the energy efficiency benefits of in-tegrating VLC with RF-based networks in a heterogeneous wireless environment, by formulating the problem of power and bandwidth allocation for energy efficiency maximization of a heterogeneous network composed of a VLC system and an RF communication system. The latter study, instead, designs an energy efficient indoor VLC system from a radically new perspective based on an amorphous user-to-network association structure, and shows that the proposed amorphous cells are capable of achieving a much higher energy efficiency per user compared to that of the conventional cell formation, for a range of practical field-of-view angles.

The paper [109] focuses on the energy efficiency aspects of local caching, and derives an approximate expression for the energy efficiency of a cache-enabled wireless network. The study shows that caching at the BSs can improve the network energy efficiency when power efficient cache hardware is used; additionally, the caching energy efficiency gain is large when the backhaul capacity is stringent, the interference level is low, content popularity is skewed, and when caching takes place at pico BSs instead of macro BSs.

Finally, the paper [110] investigates the use of mmWaves for wearable personal networks in crowded environments. An energy efficiency assessment of mmWave-based "high-end" wearables that employ advanced antenna beamforming techniques is proposed; first analytical results on the underlying scaling laws for the interacting mmWave-based networks based on IEEE 802.11ad are given, and then the impact of beamforming quality on the system energy efficiency under various conditions is quantified.

## IV. Energy harvesting and Transfer

Harvesting energy from the environment and converting it to electrical power is emerging as an appealing possibility to operate wireless communication systems. Indeed, although this approach does not directly reduce the amount of energy required to operate the system, it enables wireless networks to be powered by renewable and clean energy sources [111]. Two main kinds of energy harvesting have emerged so far in the context of wireless communications.

- **Environmental energy harvesting.** This technique refers to harvesting clean energy from natural sources, such as sun and wind. Comprehensive surveys on this approach are [112] and [113].

- **Radio-frequency energy harvesting.** This technique refers to harvesting energy from the radio signals over the air, thus enabling the recycling of energy that would otherwise be wasted. In this context, interference signals provide a natural source of electromagnetic-based power. Surveys on this approach are [114] and [115].

The main challenge in the design of communication systems powered by energy harvesting is the random amount of energy available at any given time. This is due to the fact that the availability of environmental energy sources (e.g. sun or wind) is inherently a stochastic process, and poses the problem of energy outages. Unlike traditionally-powered networks, communication systems powered by energy harvesting must comply with the so-called *energy causality* constraint, i.e. the energy used at time $t$ cannot exceed the energy harvested up to time $t$.

Early works on environmental energy harvesting dealt with this problem by taking a so-called *off-line* approach, assuming that the amount of energy harvested at a given point in time is known in advance. Although difficult to meet in practice, this approach provides insight as to the ultimate performance of energy-harvesting systems. In [116] an offline power allocation algorithm termed directional waterfilling is proposed, while [117] addresses a similar problem but assuming a system in which the data to be transmitted is available at random times. In [118] and [119], the results of [117] are extended to the more realistic case of a battery with finite capacity, while the impact of energy leakages due to non-ideal batteries is considered in [120]. Previous results have been extended to multi-user networks in [121] and [122], to relay-assisted communications in [123], and to multiple-antenna systems [124].

More recently, research efforts have been aimed at overcoming the off-line approach, developing *on-line* design policies, which do not assume any knowledge about the amount of energy harvested at specific times. Two main approaches have emerged in this context. Tools from *stochastic optimization* are used to develop design protocols assuming that the statistics of the energy process are known [125]–[127]. Alternatively, approaches based on *learning theory* provide the means to design energy harvesting systems by having the users adapt to the environment based on past observations [128], [129].

The issue of energy randomness is also present as far as radio-frequency energy harvesting is concerned, because in general the amount of electromagnetic power available in the air is not known in advance. Indeed, several schemes have appeared in the literature in which a node opportunistically exploits the electromagnetic radiation over the air. In [130] an OFDMA system is considered, in which a hybrid BS is considered, which is partly powered by radio frequency energy harvesting. In [131] and [132] a relay-assisted network is considered, wherein the relay is powered by drawing power from the received signals. A cognitive radio system is considered in [133], in which the secondary network draws energy from the signals received from the primary network.

However, radio-frequency energy harvesting offers an intriguing possibility, which also helps to reduce the randomness of wireless power sources. The idea is to combine energy harvesting with *wireless power transfer* techniques, thereby enabling network nodes to share energy with one another [134]. This has a two-fold advantage. First, it makes it possible to redistribute the network total energy, prolonging the lifetime of nodes that are low on battery energy [135], [136]. Second, it is possible to deploy dedicated beacons in the network, which act as wireless energy sources, thereby eliminating or reducing the randomness of the radio-frequency energy source. This approach can be taken even further, superimposing the energy signals on regular communication signals, resulting in the so-called *simultaneous wireless information and power transfer* (SWIPT) [137]–[139].

Several contributions to wireless power transfer are included in this special issue [140]–[142]. In [140], SWIPT in non-orthogonal multiple access networks is considered. The network nodes are assumed to be spatially randomly located over the covered area and a novel protocol is provided in which users close to the source act as energy harvesting relays to help faraway users. In [141], the co-existence of a MISO femtocell system with a macro-cell system is considered. The femtocell simultaneously transmits information to some of its users and energy to the rest of its users, while also suppressing its interference to macro-cell devices. The system energy efficiency is maximized with respect to the system beamforming vectors by means of fractional programming theory. In [142], energy harvesting and wireless power transfer is studied in relay-assisted systems with distributed beamforming, proposing a novel power splitting strategy.

## V. Hardware solutions

Energy-efficient hardware solutions refers to a broad category of strategies comprising the green design of the RF chain, the use of simplified transmitter/receiver structures, and, also, a novel architectural design of the network based on a cloud implementation of the radio access network (RAN) and on the use of network function virtualization.

Attention has been given to the energy-efficient design of power amplifiers [143], [144], both through direct circuit design and through signal design techniques aimed at peak-to-average-power ratio reduction. The use of simplified transmitter and receiver architectures, including the adoption of coarse signal quantization (e.g. one bit quantization) and hybrid analog/digital beamformers, is another technique that is being proposed for increasing hardware energy efficiency, especially in systems with many antennas such as massive MIMO systems and mmWave systems. The paper [145], as an instance, presents an analysis of the spectral efficiency of single-carrier and OFDM transmission in massive MIMO systems that use one-bit analog-to-digital converters (ADCs), while a capacity analysis of one-bit quantized MIMO systems with transmitter CSI is reported in [146]. One-bit ADCs coupled with high-resolution ADCs are instead proposed and analyzed in the paper [147], from this special issue, to simplify receiver design in massive MIMO systems. The paper shows that the proposed mixed-ADC architecture with a relatively small number of high-resolution ADCs is able to achieve a large fraction of the channel capacity of the conventional architecture, while reducing the energy consumption considerably even compared with antenna selection strategies, for both single-user and multi-user scenarios.

For mmWave communications, given the required large number of antenna elements, the implementation of digital beamforming poses serious complexity, energy consumption, and cost issues. Hybrid analog and digital beamforming structures have been thus proposed as a viable approach to reduce complexity and, most relevant to us, energy consumption [104], [148], [149]. The paper [150], in this special issue, focuses on a mmWave MIMO link with hybrid decoding. Unlike previous contributions on the subject, which considered a fully-connected architecture requiring a large number of phase shifters, a more energy-efficient hybrid precoding with sub-connected architecture is proposed and analyzed in conjunction with a successive interference cancellation (SIC) strategy. The paper also shows through simulation results that the proposed SIC-based hybrid precoding is near-optimal and enjoys higher energy efficiency than spatially sparse precoding [151] and fully digital precoding.

Cloud-based implementation of the RAN is another key technology instrumental to making future 5G networks more energy-efficient. Spurred by the impressive spread of cloud computing, cloud-RAN (C-RAN) is based on the idea that many functions that are currently performed in the BS, can be actually transferred to a remote data-center and implemented via software [17], [152], [153]. The most extreme implementation of C-RAN foresees light BSs wherein only the RF chain and the baseband-to-RF conversion stages are present; it is assumed that these light BSs are connected through high-capacity links to the data-center, wherein all the baseband processing and the resource allocation algorithms are run. This enables a great deal of flexibility in the network, thus leading to substantial savings as far as both deployment costs and energy consumption are concerned. Mobile-edge computing [154] is also a recently considered approach that increases network flexibility potentially leading to considerable energy savings. The studies [155]–[158] are a sample of the many recent works that have addressed the energy-efficiency gains possible with a cloud-based RAN.

In this special issue, paper [159] investigates the role that cellular traffic dynamics play in efficient network energy management, and designs a framework for traffic-aware energy optimization. In particular, using a learning approach, it is shown that the C-RAN can be made aware of the near-future traffic, so that inactive or low-load BSs can be switched off, thus reducing the overall energy consumption. The proposed approach is also validated on real traffic traces and energy savings on the order of 25% are achieved. The paper [160], from this special issue, proposes a holistic sparse optimization framework to design a green C-RAN by taking into consideration the power consumption of the fronthaul links, multicast services, as well as user admission control. Specifically, the sparsity structures in the solutions of both the network power minimization and user admission control problems are identified, which call for adaptive remote radio head (RRH) selection and user admission, a problem that is solved through a nonconvex but smoothed $\ell_p$-minimization ($0 < p \leq 1$) approach to promote sparsity in the multicast setting. Finally, [161], again from this special issue, studies the energy efficiency of a downlink C-RAN, focusing on two different downlink transmission strategies, namely the data-sharing strategy and the compression strategy. The paper shows that C-RAN signicantly improves the range of feasible user data rates in a wireless cellular network, and that both data-sharing and compression strategies bring much improved energy efficiency to downlink C-RAN as compared to non-optimized Coordinated Multipoint (CoMP).

## VI. FUTURE RESEARCH CHALLENGES

After having reviewed the state-of-the-art of the main 5G energy-efficient techniques, a natural question is: what are the next steps to be taken towards an energy-efficient 5G? We review some of them in the following.

### A. The need for a holistic approach

In our opinion, the main issue concerning the current state-of-the art is that most research has been directed towards a separate analysis and use of the different energy-efficient technologies. Resource allocation, deployment and planning methods, energy harvesting and transfer, have been mostly studied separately, but will a single approach be able to achieve the desired thousand-fold energy efficiency increase with respect to present networks? Most likely the answer is negative. A holistic approach is thus necessary, in which all energy-efficient techniques are combined. Indeed, as previously discussed, some works in this special issue go in this direction combining multiple energy-efficient techniques together. The GreenTouch project [8], [9] has taken an initial end-to-end perspective for the assessment of the network energy efficiency and energy consumption. More research in this direction is needed to understand the relative impact and the combined benefits of new technologies, architectures and algorithms being developed.

### B. Dealing with interference

Section II has introduced fractional programming as the most suitable tool to handle energy-efficient resource allocation. However, direct application of fractional programming typically requires a prohibitive complexity to operate in interference-limited networks. Unfortunately, 5G networks will be interference-limited, since orthogonal transmission schemes and/or linear interference neutralization techniques are not practical due to the massive amount of nodes to be served. Thus, the potentialities of fractional programming must be extended. A promising answer is represented by the framework of *sequential fractional programming*, which provides a systematic approach to extend fractional programming to interference-limited networks with affordable complexity. Sequential fractional programming has been recently shown to be effective in optimizing the energy efficiency of a number of candidate technologies for 5G, such as C-RAN, CoMP, and multi-cell systems, also with multi-carrier transmissions [23], [47], multi-cell massive MIMO systems [47], heterogeneous relay-assisted interference networks [46], full-duplex systems [45], and device-to-device systems [94], [162].

### C. Dealing with randomness

Previous sections have shown how randomness will be a pervasive feature of future wireless communication systems, which will affect network topologies, traffic evolution, and energy availability. The energy-efficient design of networks with such an unprecedented level of randomness requires the development of new statistical models, which are able to capture the average or limiting behavior of randomly evolving networks. Random matrix theory and stochastic geometry appear as suitable tools towards this end, but most studies employing these techniques have been concerned with traditional performance metrics, whereas a thorough investigation of their impact on the energy efficiency of communication systems is still missing.

A second approach lies in the use of learning techniques, which deal with randomness by letting the devices learn from past observations of their surroundings and respond as appropriate in a self-organizing fashion. However, also in this case, very little research effort has been directed towards understanding the impact of this technique on energy-efficient network design.

### D. Emerging techniques and new energy models

In addition, new emerging technologies can also be used for energy-efficient purposes. In particular, caching and mobile computing have shown significant potential as far as reducing energy consumption is concerned. By an intelligent distribution of frequently accessed content over the network nodes, caching alleviates the need for backhaul transmissions, which results in relevant energy consumption reductions. Instead, mobile computing does not directly reduce the energy consumption, but, similarly to wireless power transfer, it can prolong the lifetime of nodes that are low on battery energy. Nevertheless, in order to conclusively quantify the impact of these techniques on energy efficiency it is necessary to develop new energy consumption models which take into account the energy consumption associated with overhead transmissions over the backhaul, to feedback signaling, and to the execution of computing operations in digital signal processors.

## VII. Conclusions

Wireless communications are undergoing a rapid evolution, wherein the quest for new services and applications pushes for the fast introduction of new technologies into the marketplace. Operators are just now starting to make initial profits from their deployed LTE networks, and already 5G demos and prototypes are being announced. Moreover, the wireless communications industry has begun to design for energy efficiency. As shown in this survey, energy efficiency has gained in the last decade its own role as a performance measure and design constraint for communication networks, but many technical, regulatory, policy, and business challenges still remain to be addressed before the ambitious 1000-times energy efficiency improvement goal can be reached. We hope that this paper and those in this special issue will help to move us forward along this road.

## References

[1] Ericsson White Paper, "More than 50 billion connected devices," Ericsson, Tech. Rep. 284 23-3149 Uen, Feb. 2011.

[2] "The 1000x data challenge," Qualcomm, Tech. Rep. [Online]. Available: http://www.qualcomm.com/1000x

[3] A. Fehske, J. Malmodin, G. Biczók, and G. Fettweis, "The Global Footprint of Mobile Communications–The Ecological and Economic Perspective," *IEEE Communications Magazine, issue on Green Communications*, pp. 55–62, August 2011.

[4] G. Auer, V. Giannini, C. Desset, I. Godor, P. Skillermark, M. Olsson, M. A. Imran, D. Sabella, M. J. Gonzalez, O. Blume, and A. Fehske, "How much energy is needed to run a wireless network?" *IEEE Wireless Communications*, vol. 18, no. 5, pp. 40–49, 2011.

[5] "Why the EU is betting big on 5G," *Research EU Focus Magazine*, vol. 15, 2015.

[6] "NGMM alliance 5G white paper," *https://www.ngmm.org/5g-white-paper/5g-white-paper.html*, 2015.

[7] J. G. Andrews, S. Buzzi, W. Choi, S. Hanly, A. Lozano, A. C. K. Soong, and J. C. Zhang, "What will 5G be?" *IEEE Journal on Selected Areas in Communications*, vol. 32, no. 6, pp. 1065–1082, June 2014.

[8] "The GreenTouch Project," http://www.greentouch.org, accessed: 2016-03-22.

[9] GreenTouch Foundation, "Reducing the net energy consumption in communications networks by up to 98% by 2020," Tech. Rep., 2015.

[10] D. J. Goodman and N. B. Mandayam, "Power control for wireless data," *IEEE Personal Communications*, vol. 7, pp. 48–54, 2000.

[11] C. U. Saraydar, N. B. Mandayam, and D. J. Goodman, "Pricing and power control in a multicell wireless data network," *IEEE Journal on Selected Areas in Communications*, vol. 19, no. 10, pp. 1883–1892, October 2001.

[12] F. Meshkati, H. V. Poor, S. C. Schwartz, and N. B. Mandayam, "An energy-efficient approach to power control and receiver design in wireless data networks," *IEEE Transactions on Communications*, vol. 53, no. 11, pp. 1885–1894, November 2005.

[13] A. Zappone and E. Jorswieck, "Energy efficiency in wireless networks via fractional programming theory," *Foundations and Trends in Communications and Information Theory*, vol. 11, no. 3-4, pp. 185–396, 2015.

[14] Z. Niu, Y. Wu, J. Gong, and Z. Yang, "Cell zooming for cost-efficient green cellular networks," *IEEE Communications Magazine*, vol. 48, no. 11, pp. 74–79, November 2010.

[15] E. Oh, K. Son, and B. Krishnamachari, "Dynamic base station switching-on/off strategies for green cellular networks," *IEEE Transactions on Wireless Communications*, vol. 12, no. 5, pp. 2126–2136, 2013.

[16] C. Han, T. Harrold, S. Armour, I. Krikidis, S. Videv, P. M. Grant, H. Haas, J. S. Thompson, I. Ku, C.-X. Wang *et al.*, "Green radio: radio techniques to enable energy-efficient wireless networks," *IEEE Communications Magazine*, vol. 49, no. 6, pp. 46–54, 2011.

[17] P. Rost, C. Bernardos, A. Domenico, M. Girolamo, M. Lalam, A. Maeder, D. Sabella, and D. Wübben, "Cloud technologies for flexible 5G radio access networks," *IEEE Communications Magazine*, vol. 52, no. 5, pp. 68–76, 2014.

[18] C. Isheden, Z. Chong, E. A. Jorswieck, and G. Fettweis, "Framework for link-level energy efficiency optimization with informed transmitter," *IEEE Transactions on Wireless Communications*, vol. 11, no. 8, pp. 2946–2957, August 2012.

[19] F. R. Yu, X. Zhang, and V. C. Leung, *Green Communications and Networking*. CRC Press, 2012.

[20] E. Hossain, V. K. Bhargava, and G. Fettweis, Eds., *Green Radio Communication Networks*. Cambridge University Press, 2012.

[21] G. Miao, N. Himayat, G. Y. Li, and S. Talwar, "Distributed interference-aware energy-efficient power optimization," *IEEE Transactions on Wireless Communications*, vol. 10, no. 4, pp. 1323–1333, April 2011.

[22] D. W. K. Ng, E. S. Lo, and R. Schober, "Energy-efficient resource allocation in multi-cell OFDMA systems with limited backhaul capacity," *IEEE Transactions on Wireless Communications*, vol. 11, no. 10, pp. 3618–3631, October 2012.

[23] L. Venturino, A. Zappone, C. Risi, and S. Buzzi, "Energy-efficient scheduling and power allocation in downlink OFDMA networks with base station coordination," *IEEE Transactions on Wireless Communications*, vol. 14, no. 1, pp. 1–14, January 2015.

[24] J. Xu and L. Qiu, "Energy efficiency optimization for MIMO broadcast channels," *IEEE Transactions on Wireless Communications*, vol. 12, no. 2, pp. 690–701, February 2013.

[25] A. Zappone, P. Cao, and E. A. Jorswieck, "Energy efficiency optimization in relay-assisted MIMO systems with perfect and statistical CSI," *IEEE Transactions on Signal Processing*, vol. 62, no. 2, pp. 443–457, January 2014.

[26] O. Onireti, F. Heliot, and M. A. Imran, "On the energy efficiency-spectral efficiency trade-off of distributed MIMO systems," *IEEE Transactions on Communications*, vol. 61, no. 9, pp. 3741–3753, September 2013.

[27] K. T. K. Cheung, S. Yang, and L. Hanzo, "Achieving maximum energy-efficiency in multi-relay OFDMA cellular networks: A fractional programming approach," *IEEE Transactions on Communications*, vol. 61, no. 7, pp. 2746–2757, 2013.

[28] C. Sun and C. Yang, "Energy efficiency analysis of one-way and two-way relay systems," *EURASIP Journal on Wireless Communications and Networking*, pp. 1–18, February 2012.

[29] B. Matthiesen, A. Zappone, and E. A. Jorswieck, "Design of 3-way relay channels for throughput and energy efficiency," *IEEE Transactions on Wireless Communications*, vol. 14, no. 8, pp. 4454–4468, August 2014.

[30] Y. Wang, W. Xu, K. Yang, and J. Lin, "Optimal energy-efficient power allocation for OFDM-based cognitive radio networks," *IEEE Communications Letters*, vol. 16, no. 9, pp. 1420–1423, 2012.

[31] F. Gabry, A. Zappone, R. Thobaben, E. A. Jorswieck, and M. Skoglund, "Energy efficiency analysis of cooperative jamming in cognitive radio networks with secrecy constraints," *IEEE Wireless Communications Letters*, vol. 4, no. 4, pp. 437–440, August 2015.

[32] F. Meshkati, M. Chiang, H. V. Poor, and S. C. Schwartz, "A game-theoretic approach to energy-efficient power control in multicarrier CDMA systems," *IEEE Journal on Selected Areas in Communications*, vol. 24, no. 6, pp. 1115–1129, June 2006.

[33] S. Buzzi and H. V. Poor, "Joint receiver and transmitter optimization for energy-efficient CDMA communications," *IEEE Journal on Selected Areas in Communications*, vol. 26, no. 3, pp. 459–472, April 2008.

[34] S. M. Betz and H. V. Poor, "Energy efficient communications in CDMA networks: A game theoretic analysis considering operating costs," *IEEE Transactions on Signal Processing*, vol. 56, no. 10, pp. 5181–5190, October 2008.

[35] S. Lasaulce, Y. Hayel, R. El Azouzi, and M. Debbah, "Introducing hierarchy in energy games," *IEEE Transactions on Wireless Communications*, vol. 8, no. 7, pp. 3833–3843, 2009.

[36] M. L. Treust and S. Lasaulce, "A repeated game formulation of energy-efficient decentralized power control," *IEEE Transactions on Wireless Communications*, vol. 9, no. 9, pp. 2860–2869, September 2010.

[37] S. Buzzi and D. Saturnino, "A game-theoretic approach to energy-efficient power control and receiver design in cognitive CDMA wireless networks," *IEEE Journal of Selected Topics in Signal Processing*, vol. 5, no. 1, pp. 137–150, February 2011.

[38] S. Buzzi, H. V. Poor, and A. Zappone, "Transmitter waveform and widely-linear receiver design: Non-cooperative games for wireless multiple-access networks," *IEEE Transactions on Information Theory*, vol. 56, no. 10, pp. 4874–4892, October 2010.

[39] G. Bacci, H. V. Poor, M. Luise, and A. M. Tulino, "Energy efficient power control in impulse radio UWB wireless networks," *IEEE Journal of Selected Topics in Signal Processing*, vol. 1, no. 3, pp. 508–520, October 2007.

[40] A. Zappone, Z. Chong, E. A. Jorswieck, and S. Buzzi, "Energy-aware competitive power control in relay-assisted interference wireless networks," *IEEE Transactions on Wireless Communications*, vol. 12, no. 4, pp. 1860–1871, April 2013.

[41] Z. Xu, C. Yang, G. Y. Li, S. Zhang, Y. Chen, and S. Xu, "Energy-efficient configuration of spatial and frequency resources in MIMO-OFDMA systems," *IEEE Transactions on Communications*, vol. 61, no. 2, pp. 564–575, 2013.

[42] E. V. Belmega and S. Lasaulce, "Energy-efficient precoding for multiple-antenna terminals," *IEEE Transactions on Signal Processing*, vol. 59, no. 1, pp. 329–340, January 2011.

[43] G. Brante, I. Stupia, R. D. Souza, and L. Vandendorpe, "Outage probability and energy efficiency of cooperative MIMO with antenna selection," *IEEE Transactions on Wireless Communications*, vol. 12, no. 11, pp. 5896–5907, November 2013.

[44] C. He, G. Y. Li, F.-C. Zheng, and X. You, "Energy-efficient resource allocation in OFDM systems with distributed antennas," *IEEE Transactions on Vehicular Technology*, vol. 63, no. 3, pp. 1223–1231, March 2014.

[45] D. Nguyen, L.-N. Tran, P. Pirinen, and M. Latva-aho, "Precoding for full duplex multiuser MIMO systems: Spectral and energy efficiency maximization," *IEEE Transactions on Signal Processing*, vol. 61, no. 16, pp. 4038–4050, August 2013.

[46] A. Zappone, E. A. Jorswieck, and S. Buzzi, "Energy efficiency and interference neutralization in two-hop MIMO interference channels," *IEEE Transactions on Signal Processing*, vol. 62, no. 24, pp. 6481–6495, December 2014.

[47] A. Zappone, S. Sanguinetti, G. Bacci, E. A. Jorswieck, and M. Debbah, "Energy-efficient power control: A look at 5G wireless technologies," *IEEE Transactions on Signal Processing*, vol. 64, no. 7, pp. 1668–1683, April 2016.

[48] L. Venturino and S. Buzzi, "Energy-aware and rate-aware heuristic beamforming in downlink MIMO OFDMA networks with base-station coordination," *IEEE Transactions on Vehicular Technology*, vol. 64, no. 7, pp. 2897–2910, 2015.

[49] S. He, Y. Huang, L. Yang, and B. Ottersten, "Coordinated multicell multiuser precoding for maximizing weighted sum energy efficiency,"

[50] S. Buzzi, G. Colavolpe, D. Saturnino, and A. Zappone, "Potential games for energy-efficient power control and subcarrier allocation in uplink multicell OFDMA systems," *IEEE Journal of Selected Topics in Signal Processing*, vol. 6, no. 2, pp. 89 –103, April 2012.

[51] B. Du, C. Pan, W. Zhang, and M. Chen, "Distributed energy-efficient power optimization for CoMP systems with max-min fairness," *IEEE Communications Letters*, vol. 18, no. 6, pp. 999–1002, 2014.

[52] G. Bacci, E. V. Belmega, P. Mertikopoulos, and L. Sanguinetti, "Energy-aware competitive power allocation in heterogeneous networks with QoS constraints," *IEEE Transactions on Wireless Communications*, vol. 14, no. 9, pp. 4728–4742, September 2015.

[53] F. Meshkati, A. J. Goldsmith, H. V. Poor, and S. C. Schwartz, "A game-theoretic approach to energy-efficient modulation in CDMA networks with delay QoS constraints," *IEEE Journal on Selected Areas in Communications*, vol. 25, no. 6, pp. 1069–1078, August 2007.

[54] M. Sinaie, A. Zappone, E. Jorswieck, and P. Azmi, "A novel power consumption model for effective energy efficiency in wireless networks," *IEEE Wireless Communications Letters*, vol. PP, no. 99, 2016.

[55] C. She, C. Yang, and L. Liu, "Energy-efficient resource allocation for MIMO-OFDM systems serving random sources with statistical QoS requirement," *IEEE Transactions on Communications*, vol. 63, no. 11, pp. 4125–4141, 2015.

[56] J. Lv, A. Zappone, and E. A. Jorswieck, "Energy-efficient MIMO underlay spectrum sharing with rate splitting," in *Proc. 15th IEEE International Workshop on Signal Processing Advances in Wireless Communications (SPAWC 14)*, Toronto, Canada, June 2014.

[57] D. López-Pérez, X. Chu, A. V. Vasilakos, and H. Claussen, "Power minimization based resource allocation for interference mitigation in OFDMA femtocell networks," *IEEE Journal on Selected Areas Communications*, vol. 32, no. 2, pp. 333–344, February 2014.

[58] M. Moretti, L. Sanguinetti, and X. Wang, "Resource allocation for power minimization in the downlink of THP-based spatial multiplexing MIMO-OFDMA systems," *IEEE Transactions on Vehicular Technology*, vol. 64, no. 1, pp. 405–411, January 2015.

[59] C. Pan, W. Xu, J. Wang, H. Ren, W. Zhang, N. Huang, and M. Chen, "Pricing-based distributed energy-efficient beamforming for MISO interference channels," *IEEE Journal on Selected Areas in Communications*, vol. 34, no. 4, April 2016.

[60] Q. Chen, G. Yu, R. Yin, A. Maaref, G. Y. Li, and A. Huang, "Energy efficiency optimization in licensed-assisted access," *IEEE Journal on Selected Areas in Communications*, vol. 34, no. 4, April 2016.

[61] H. Yu, M. H. Cheung, L. Huang, and J. Huang, "Power-delay tradeoff with predictive scheduling in integrated cellular and Wi-Fi networks," *IEEE Journal on Selected Areas in Communications*, vol. 34, no. 4, April 2016.

[62] P. Mertikopoulos and E. V. Belmega, "Learning to be green: Robust energy efficiency maximization in dynamic MIMO-OFDM systems," *IEEE Journal on Selected Areas in Communications*, vol. 34, no. 4, April 2016.

[63] R. Zi, X. Ge, J. Thompson, C.-X. Wang, H. Wang, and T. Han, "Energy efficiency optimization of 5G radio frequency chain systems," *IEEE Journal on Selected Areas in Communications*, vol. 34, no. 4, April 2016.

[64] H. Park and T. Hwang, "Energy-efficient power control of cognitive femto users for 5G communications," *IEEE Journal on Selected Areas in Communications*, vol. 34, no. 4, April 2016.

[65] J. G. Andrews, "Seven ways that HetNets are a cellular paradigm shift," *IEEE Communications Magazine*, vol. 51, no. 3, pp. 136–144, March 2013.

[66] F. Baccelli and B. Błaszczyszyn, "Stochastic geometry and wireless networks: Volume I theory," *Now Publishers: Foundations and Trends in Networking*, vol. 3, no. 3–4, pp. 249–449, March 2009.

[67] S. Weber and J. G. Andrews, "Transmission capacity of wireless networks," *Now Publishers: Foundations and Trends in Communications and Information Theory*, vol. 5, no. 2–3, pp. 109–281, 2012.

[68] I. Guvenc, T. Q. S. Quek, M. Kountoris, and L. Perez, "Special issue on heterogeneous and small cell networks: Part 1," *IEEE Communication Magazine*, vol. 51, no. 5, May 2013.

[69] ——, "Special issue on heterogeneous and small cell networks: Part 2," *IEEE Communication Magazine*, vol. 51, no. 6, June 2013.

[70] Y. S. Soh, T. Q. S. Quek, M. Kountouris, and H. Shin, "Energy efficient heterogeneous cellular networks," *IEEE Journal on Selected Areas in Communications*, vol. 31, no. 5, pp. 840–850, May 2013.

[71] D. Cao, S. Zhou, and Z. Niu, "Optimal combination of base station densities for energy-efficient two-tier heterogeneous cellular networks,"

*IEEE Transaction on Wireless Communications*, vol. 12, no. 9, September 2013.

[72] S. Kim, B. G. Lee, and D. Park, "Energy-per-bit minimized radio resource allocation in heterogeneous networks," *IEEE Transactions on Wireless Communications*, vol. 13, no. 4, pp. 1862–1873, April 2014.

[73] J. Xu, L. Duan, and R. Zhang, "Energy group buying with loading sharing for green cellular networks," *IEEE Journal on Selected Areas in Communications*, vol. 34, no. 4, April 2016.

[74] T. L. Marzetta, "Noncooperative cellular wireless with unlimited numbers of base station antennas," *IEEE Transactions on Wireless Communications*, vol. 9, no. 11, pp. 3590–3600, November 2010.

[75] F. Rusek, D. Persson, B. K. Lau, E. G. Larsson, T. L. Marzetta, O. Edfors, and F. Tufvesson, "Scaling up MIMO: Opportunities and challenges with very large arrays," *IEEE Signal Processing Magazine*, vol. 30, no. 1, pp. 40–60, January 2013.

[76] E. G. Larsson, O. Edfors, F. Tufvesson, and T. L. Marzetta, "Massive MIMO for next generation wireless systems," *IEEE Communications Magazine*, vol. 52, no. 2, pp. 186–195, February 2014.

[77] R. Couillet and M. Debbah, *Random Matrix Methods for Wireless Communications*. Cambridge, New York: Cambridge University Press, 2011.

[78] A. M. Tulino and S. Verdú, "Random matrix theory and wireless communications," *Now Publishers: Foundations and Trends in Communications and Information Theory*, vol. 1, no. 1, pp. 1–182, June 2004.

[79] H. Yang and T. L. Marzetta, "Performance of conjugate and zero-forcing beamforming in large-scale antenna systems," *IEEE Journal on Selected Areas in Communications*, vol. 31, no. 2, February 2013.

[80] J. Hoydis, S. ten Brink, and M. Debbah, "Massive MIMO in the UL/DL of cellular networks: How many antennas do we need?" *IEEE Journal on Selected Areas in Communications*, vol. 31, no. 2, pp. 160–171, February 2013.

[81] H. Q. Ngo, E. G. Larsson, and T. L. Marzetta, "Energy and spectral efficiency of very large multiuser MIMO systems," *IEEE Transactions on Communications*, vol. 61, no. 4, April 2013.

[82] E. Björnson, J. Hoydis, M. Kountouris, and M. Debbah, "Massive MIMO systems with non-ideal hardware: Energy efficiency, estimation, and capacity limits," *IEEE Transaction on Information Theory*, vol. 60, no. 11, pp. 7112–7139, November 2014.

[83] E. Björnson, L. Sanguinetti, J. Hoydis, and M. Debbah, "Optimal design of energy-efficient multi-user MIMO systems: Is massive MIMO the answer?" *IEEE Transactions on Wireless Communications*, vol. 14, no. 6, pp. 3059–3075, June 2015.

[84] H. Klessig, D. Ohmann, A. I. Reppas, H. Hatzikirou, M. Abedi, M. Simsek, and G. Fettweis, "From immune cells to self-organizing ultra-dense small cell networks," *IEEE Journal on Selected Areas in Communications*, vol. 34, no. 4, April 2016.

[85] M. Fiorani, S. Tombaz, F. Farias, L. Wosinska, and P. Monti, "Joint design of radio and transport for green residential access networks," *IEEE Journal on Selected Areas in Communications*, vol. 34, no. 4, April 2016.

[86] B. Zhuang, D. Guo, and M. L. Honig, "Energy-efficient cell activation, user association, and spectrum allocation in heterogeneous networks," *IEEE Journal on Selected Areas in Communications*, vol. 34, no. 4, April 2016.

[87] E. Björnson, L. Sanguinetti, and M. Kountouris, "Deploying dense networks for maximal energy efficiency: Small cells meet massive MIMO," *IEEE Journal on Selected Areas in Communications*, vol. 34, no. 4, April 2016.

[88] Z. Zhang, Z. Chen, M. Shen, and B. Xia, "Spectral and energy efficiency of multi-pair two-way full-duplex relay systems with massive MIMO," *IEEE Journal on Selected Areas in Communications*, vol. 34, no. 4, April 2016.

[89] F. Boccardi, R. W. Heath Jr., A. Lozano, T. L. Marzetta, and P. Popovski, "Five disruptive technology directions for 5G," *IEEE Communications Magazine*, vol. 52, no. 2, pp. 74–80, February 2014.

[90] M. N. Tehrani, M. Uysal, and H. Yanikomeroglu, "Device-to-device communication in 5G cellular networks: challenges, solutions, and future directions," *IEEE Communications Magazine*, vol. 52, no. 5, pp. 86–92, 2014.

[91] F. Wang, C. Xu, L. Song, Q. Zhao, X. Wang, and Z. Han, "Energy-aware resource allocation for device-to-device underlay communication," in *Proc. 2013 IEEE International Conference on Communications (ICC)*, 2013, pp. 6076–6080.

[92] D. Wu, J. Wang, R. Hu, Y. Cai, and L. Zhou, "Energy-efficient resource sharing for mobile device-to-device multimedia communica-

tions," *IEEE Transactions on Vehicular Technology*, vol. 63, no. 5, pp. 2093–2103, 2014.

[93] Z. Zhou, M. Dong, K. Ota, J. Wu, and T. Sato, "Energy efficiency and spectral efficiency tradeoff in device-to-device (D2D) communications," *IEEE Wireless Communications Letters*, vol. 3, no. 5, pp. 485–488, 2014.

[94] A. Zappone, F. D. Stasio, S. Buzzi, and E. Jorswieck, "Energy-efficient resource allocation in 5G with application to D2D," in *Signal Processing for 5G: Algorithms and Implementations*, F.-L. Luo and J. C. Zahng, Eds. Wiley, 2016, ch. 19.

[95] S. Wu, H. Wang, and C.-H. Youn, "Visible light communications for 5G wireless networking systems: From fixed to mobile communications," *IEEE Network*, vol. 28, no. 6, pp. 41–45, 2014.

[96] H. Haas, "The future of wireless light communication," in *Proc. 2nd International Workshop on Visible Light Communications Systems*. ACM, 2015, pp. 1–1.

[97] J. Kim, C. Lee, and J.-K. K. Rhee, "Traffic off-balancing algorithm for energy efficient networks," in *SPIE/OSA/IEEE Asia Communications and Photonics*. International Society for Optics and Photonics, 2011, pp. 83 100C–83 100C.

[98] M. Ji, G. Caire, and A. F. Molisch, "Wireless device-to-device caching networks: Basic principles and system performance," *IEEE Journal on Selected Areas in Communications*, vol. 34, no. 1, pp. 176–189, 2016.

[99] K. Shanmugam, N. Golrezaei, A. G. Dimakis, A. F. Molisch, and G. Caire, "Femtocaching: Wireless content delivery through distributed caching helpers," *IEEE Transactions on Information Theory*, vol. 59, no. 12, pp. 8402–8413, 2013.

[100] A. F. Molisch, G. Caire, D. Ott, J. R. Foerster, D. Bethanabhotla, and M. Ji, "Caching eliminates the wireless bottleneck in video aware wireless networks," *Advances in Electrical Engineering*, vol. 2014, 2014.

[101] E. Bastug, M. Bennis, and M. Debbah, "Living on the edge: The role of proactive caching in 5G wireless networks," *IEEE Communications Magazine*, vol. 52, no. 8, pp. 82–89, 2014.

[102] T. S. Rappaport, S. Sun, R. Mayzus, H. Zhao, Y. Azar, K. Wang, G. N. Wong, J. K. Schulz, M. Samimi, and F. Gutierrez, "Millimeter wave mobile communications for 5G cellular: It will work!" *IEEE Access*, vol. 1, pp. 335–349, 2013.

[103] S. Rangan, T. Rappaport, E. Erkip, Z. Latinovic, M. R. Akdeniz, and Y. Liu, "Energy efficient methods for millimeter wave picocellular systems," in *Proc. 2013 IEEE Communication Theory Workshop*, 2013.

[104] L. Dai, X. Gao, and Z. Wang, "Energy-efficient hybrid precoding based on successive interference cancelation for millimeter-wave massive MIMO systems," in *Proc. 2015 IEEE Radio and Antenna Days of the Indian Ocean (RADIO)*, 2015, pp. 1–2.

[105] X. Gao, L. Dai, S. Han, R. W. Heath Jr *et al.*, "Energy-efficient hybrid analog and digital precoding for mmWave MIMO systems with large antenna arrays," *arXiv preprint arXiv:1507.04592*, 2015.

[106] Y. Zhao, Y. Li, H. Zhang, N. Ge, and J. Lu, "Fundamental tradeoffs on energy-aware D2D communication underlaying cellular networks: A dynamic graph approach," *IEEE Journal on Selected Areas in Communications*, vol. 34, no. 4, April 2016.

[107] M. Kashef, M. Ismail, M. Abdallah, K. Qaraqe, and E. Serpedin, "Energy efficient resource allocation for mixed RF/VLC heterogeneous wireless networks," *IEEE Journal on Selected Areas in Communications*, vol. 34, no. 4, April 2016.

[108] R. Zhang, H. Claussen, H. Haas, and L. Hanzo, "Energy efficient visible light communications relying on amorphous cells," *IEEE Journal on Selected Areas in Communications*, vol. 34, no. 4, April 2016.

[109] D. Liu and C. Yang, "Energy efficiency of downlink networks with caching at base stations," *IEEE Journal on Selected Areas in Communications*, vol. 34, no. 4, April 2016.

[110] O. Galinina, A. Pyattaev, K. Johnsson, A. Turlikov, S. Andreev, and Y. Koucheryavy, "Assessing system-level energy efficiency of mmwave-based wearable networks," *IEEE Journal on Selected Areas in Communications*, vol. 34, no. 4, April 2016.

[111] 5G-PPP, "5G manifesto," in *Mobile World Congress*, 2015.

[112] H. A. Hassan *et al.*, "Renewable energy in cellular networks: A survey," in *Proc.2013 IEEE Online Conference on Green Communications*, October 2013.

[113] S. Ulukus, A. Yener, E. Erkip, O. Simeone, M. Zorzi, P. Grover, and K. Huang, "Energy harvesting wireless communications: A review of recent advances," *IEEE Journal on Selected Areas in Communications*, vol. 33, no. 3, pp. 360–380, March 2015.

[114] X. Lu, D. Niyato, D. I. Kim, and Z. Han, "Wireless networks with RF energy harvesting: A contemporary survey," *IEEE Communications Surveys & Tutorials*, vol. 17, no. 2, pp. 757–789, 2015.

[115] H. J. Visser and R. J. M. Vullers, "RF energy harvesting and transport for wireless sensor network applications: Principles and requirements," *Proceedings of the IEEE*, vol. 101, no. 6, pp. 1410–1423, June 2013.

[116] O. Ozel, K. Tutuncuoglu, J. Yang, S. Ulukus, and A. Yener, "Transmission with energy harvesting nodes in fading wireless channels: Optimal policies," *IEEE Journal on Selected Areas Communications*, vol. 29, no. 8, pp. 1732–1743, September 2011.

[117] J. Yang and S. Ulukus, "Optimal packet scheduling in an energy harvesting communication system," *IEEE Transactions on Communications*, vol. 60, no. 1, pp. 220–230, January 2012.

[118] K. Tutuncuoglu and A. Yener, "Optimum transmission policies for battery limited energy harvesting nodes," *IEEE Transactions on Wireless Communications*, vol. 11, no. 3, pp. 1180–1189, March 2012.

[119] M. Gregori and M. Payaró, "Energy-efficient transmission for wireless energy harvesting nodes," *IEEE Transactions on Wireless Communications*, vol. 12, no. 3, pp. 1244–1254, March 2013.

[120] B. Devillers and D. Gündüz, "A general framework for the optimization of energy harvesting communication systems with battery imperfections," *Journal of Communications and Networks*, vol. 14, no. 2, pp. 130–139, April 2012.

[121] J. Yang, O. Ozel, and S. Ulukus, "Broadcasting with an energy harvesting rechargeable transmitter," *IEEE Transactions on Wireless Communications*, vol. 11, no. 2, pp. 571–583, February 2012.

[122] K. Tutuncuoglu and A. Yener, "Sum-rate optimal power policies for energy harvesting transmitters in an interference channel," *Journal of Communications and Networks*, vol. 14, no. 2, April 2012.

[123] C. Huang, R. Zhang, and S. Cui, "Throughput maximization for the Gaussian relay channel with energy harvesting constraints," *IEEE Journal on Selected Areas in Communications*, vol. 31, no. 8, pp. 1469–1479, August 2013.

[124] C. Xing, N. Wang, J. Ni, Z. Fei, and J. Kuang, "MIMO beamforming designs with partial CSI under energy harvesting constraints," *IEEE Signal Processing Letters*, vol. 20, no. 4, pp. 363–366, April 2013.

[125] C. K. Ho and R. Zhang, "Optimal energy allocation for wireless communications with energy harvesting constraints," *IEEE Transactions on Signal Processing*, vol. 60, no. 9, pp. 4808–4818, September 2012.

[126] H. Huang and V. K. N. Lau, "Decentralized delay optimal control for interference networks with limited renewable energy storage," *IEEE Transactions on Signal Processing*, vol. 60, no. 5, pp. 2552–2561, May 2012.

[127] N. Michelusi and M. Zorzi, "Optimal adaptive random multiaccess in energy harvesting wireless sensor networks," *IEEE Transactions on Communications*, vol. 63, no. 4, pp. 1355–1372, April 2015.

[128] P. Blasco, D. Gündüz, and M. Dohler, "A learning theoretic approach to energy harvesting communication system optimization," *IEEE Transactions on Wireless Communications*, vol. 12, no. 4, pp. 1872–1882, April 2013.

[129] D. Gündüz, K. S. N. Michelusi, and M. Zorzi, "Designing intelligent energy harvesting communication systems," *IEEE Communications Magazine*, pp. 210–216, January 2014.

[130] D. W. K. Ng, E. S. Lo, and R. Schober, "Energy-efficient resource allocation in OFDMA systems with hybrid energy harvesting base station," *IEEE Transactions on Wireless Communications*, vol. 12, no. 7, pp. 3412–3427, July 2013.

[131] A. A. Nasir, X. Zhou, S. Durrani, and R. A. Kennedy, "Relaying protocols for wireless energy harvesting and information processing," *IEEE Transactions on Communications*, vol. 12, no. 7, pp. 3622–3636, July 2013.

[132] Z. Ding, S. M. Perlaza, I. Esnaola, and H. V. Poor, "Power allocation strategies in energy harvesting wireless cooperative networks," *IEEE Transactions on Wireless Communications*, vol. 13, no. 2, pp. 846–860, February 2014.

[133] S. Lee, R. Zhang, and K. Huang, "Opportunistic wireless energy harvesting in cognitive radio networks," *IEEE Transactions on Wireless Communications*, vol. 12, no. 9, pp. 4788–4799, September 2013.

[134] L. Liu, R. Zhang, and K.-C. Chua, "Wireless information transfer with opportunistic power harvesting," *IEEE Transactions on Wireless Communications*, vol. 12, no. 1, pp. 288–300, January 2013.

[135] B. Gurakan, O. Ozel, J. Yang, and S. Ulukus, "Energy cooperation in energy harvesting communications," *IEEE Transactions on Communications*, vol. 61, no. 12, pp. 4884–4897, December 2013.

[136] Y.-K. Chia, S. Sun, and R. Zhang, "Energy cooperation in cellular networks with renewable powered base stations," *IEEE Transactions on Wireless Communications*, vol. 13, no. 12, pp. 6996–7010, December 2014.

[137] K. Huang and E. Larsson, "Simultaneous information and power transfer for broadband wireless systems," *IEEE Transactions on Signal Processing*, vol. 61, no. 23, pp. 5972–5986, December 2013.

[138] I. Krikidis, S. Timotheou, S. Nikolaou, G. Zheng, D. W. K. Ng, and R. Schober, "Simultaneous wireless information and power transfer in modern communication systems," *IEEE Communications Magazine*, vol. 52, no. 11, pp. 104–110, November 2014.

[139] D. W. K. Ng, E. S. Lo, and R. Schober, "Wireless information and power transfer: Energy efficiency optimization in OFDMA systems," *IEEE Transactions on Wireless Communications*, vol. 12, no. 12, pp. 6352–6370, December 2013.

[140] Y. Liu, Z. Ding, M. Elkashlan, and H. V. Poor, "Cooperative non-orthogonal multiple access with simultaneous wireless information and power transfer," *IEEE Journal on Selected Areas in Communications*, vol. 34, no. 4, April 2016.

[141] M. Sheng, L. Wang, X. Wang, Y. Zhang, C. Xu, and J. Li, "Energy efficient beamforming in MISO heterogeneous cellular networks with wireless information and power transfer," *IEEE Journal on Selected Areas in Communications*, vol. 34, no. 4, April 2016.

[142] Z. Zhou, M. Peng, Z. Zhao, W. Wang, and R. S. Blum, "Wireless-powered cooperative communications: Power-splitting relaying with energy accumulation," *IEEE Journal on Selected Areas in Communications*, vol. 34, no. 4, April 2016.

[143] J. Joung, C. K. Ho, K. Adachi, and S. Sun, "A survey on power-amplifier-centric techniques for spectrum and energy-efficient wireless communications," *IEEE Communications Surveys & Tutorials*, vol. 17, no. 1, pp. 315–333, 2015.

[144] F. Mahmood, E. Perrins, and L. Liu, "Modeling and analysis of energy consumption for RF transceivers in wireless cellular systems," in *Proc. 2015 IEEE Global Communications Conference (GLOBECOM)*, 2015, pp. 1–6.

[145] C. Mollén, J. Choi, E. G. Larsson, and R. W. Heath Jr, "Performance of the wideband massive uplink MIMO with one-bit ADCs," *arXiv preprint arXiv:1602.07364*, 2016.

[146] J. Mo and R. W. Heath Jr., "Capacity analysis of one-bit quantized MIMO systems with transmitter channel state information," *IEEE Transactions on Signal Processing*, vol. 63, no. 20, pp. 5498–5512, 2015.

[147] N. Liang and W. Zhang, "Mixed-ADC massive MIMO," *IEEE Journal on Selected Areas in Communications*, vol. 34, no. 4, April 2016.

[148] S. Han, I. Chih-Lin, Z. Xu, and C. Rowell, "Large-scale antenna systems with hybrid analog and digital beamforming for millimeter wave 5G," *IEEE Communications Magazine*, vol. 53, no. 1, pp. 186–194, 2015.

[149] S. Han, C. Lin, C. Rowell, Z. Xu, S. Wang, Z. Pan *et al.*, "Large scale antenna system with hybrid digital and analog beamforming structure," in *Proc. 2014 IEEE International Conference on Communications Workshops (ICC)*, 2014, pp. 842–847.

[150] X. Gao, L. Dai, S. Han, C.-L. I, and R. W. Heath Jr., "Energy-efficient hybrid analog and digital precoding for mmwave MIMO systems with large antenna arrays," *IEEE Journal on Selected Areas in Communications*, vol. 34, no. 4, April 2016.

[151] O. El Ayach, S. Rajagopal, S. Abu-Surra, Z. Pi, and R. W. Heath Jr., "Spatially sparse precoding in millimeter wave MIMO systems," *IEEE Transactions on Wireless Communications*, vol. 13, no. 3, pp. 1499–1513, 2014.

[152] A. Checko, H. L. Christiansen, Y. Yan, L. Scolari, G. Kardaras, M. S. Berger, and L. Dittmann, "Cloud RAN for mobile networks - a technology overview," *IEEE Communications Surveys & Tutorials*, vol. 17, no. 1, pp. 405–426, 2015.

[153] P. Demestichas, A. Georgakopoulos, D. Karvounas, K. Tsagkaris, V. Stavroulaki, J. Lu, C. Xiong, and J. Yao, "5G on the horizon: key challenges for the radio-access network," *IEEE Vehicular Technology Magazine*, vol. 8, no. 3, pp. 47–53, 2013.

[154] S. Sardellitti, G. Scutari, and S. Barbarossa, "Joint optimization of radio and computational resources for multicell mobile-edge computing," *IEEE Transactions on Signal and Information Processing over Networks*, vol. 1, no. 2, pp. 89–103, 2015.

[155] D. Sabella, A. De Domenico, E. Katranaras, M. A. Imran, M. Di Girolamo, U. Salim, M. Lalam, K. Samdanis, and A. Maeder, "Energy efficiency benefits of RAN-as-a-service concept for a cloud-based 5G mobile network infrastructure," *IEEE Access*, vol. 2, pp. 1586–1597, 2014.

[156] M. Peng, Y. Li, J. Jiang, J. Li, and C. Wang, "Heterogeneous cloud radio access networks: a new perspective for enhancing spectral and energy efficiencies," *IEEE Wireless Communications*, vol. 21, no. 6, pp. 126–135, 2014.

[157] D. Pompili, A. Hajisami, and T. X. Tran, "Elastic resource utilization framework for high capacity and energy efficiency in cloud RAN," *IEEE Communications Magazine*, vol. 54, no. 1, pp. 26–32, 2016.

[158] C.-L. I, C. Rowell, S. Han, Z. Xu, G. Li, and Z. Pan, "Toward green and soft: a 5G perspective," *IEEE Communications Magazine*, vol. 52, no. 2, pp. 66–73, February 2014.

[159] N. Saxena, A. Roy, and H. Kim, "Traffic-aware cloud RAN: a key for green 5G networks," *IEEE Journal on Selected Areas in Communications*, vol. 34, no. 4, April 2016.

[160] Y. Shi, J. Cheng, J. Zhang, B. Bai, W. Chen, and K. B. Letaief, "Smoothed Lp-minimization for green cloud-RAN with user admission control," *IEEE Journal on Selected Areas in Communications*, vol. 34, no. 4, April 2016.

[161] B. Dai and W. Yu, "Energy efficiency of downlink transmission strategies for cloud radio access networks," *IEEE Journal on Selected Areas in Communications*, vol. 34, no. 4, April 2016.

[162] A. Zappone, B. Matthiesen, and E. A. Jorswieck, "Energy efficiency in MIMO underlay and overlay device-to-device communications and cognitive radio systems," *IEEE Transactions on Signal Processing, submitted, http://arxiv.org/abs/1509.08309, 2016.*

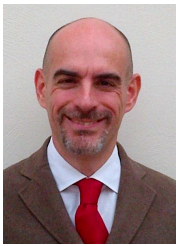

**Stefano Buzzi** (SM'07) is currently an Associate Professor at the University of Cassino and Lazio Meridionale, Italy. He received his Ph.D. degree in Electronic Engineering and Computer Science from the University of Naples "Federico II" in 1999, and he has had short-term visiting appointments at the Dept. of Electrical Engineering, Princeton University, in 1999, 2000, 2001 and 2006. His research and study interest lie in the wide area of statistical signal processing and resource allocation for communications, with emphasis on wireless cellular communications.

Dr. Buzzi is author/co-author of about 60 journal papers and 90 conference papers; he is currently Associate Editor for the *IEEE Transactions on Wireless Communications*, and a former Associate Editor for the *IEEE Communications Letters*, and the *IEEE Signal Processing Letters*. Prof. Buzzi has recently been the lead guest editor for the special issue on "5G Wireless Communication Systems," *IEEE Journal on Selected Areas in Communications*, June 2014.

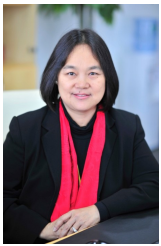

**Chih-Lin I** received her Ph.D. degree in electrical engineering from Stanford University. She has been working at multiple world-class companies and research institutes leading the R&D, including AT&T Bell Labs; Director of AT&T HQ, Director of ITRI Taiwan, and VPGD of ASTRI Hong Kong. She received the *IEEE Trans. COM* Stephen Rice Best Paper Award, is a winner of the CCCP National 1000 Talent Program, and has won the 2015 Industrial Innovation Award of IEEE Communication Society for Leadership and Innovation in Next-Generation Cellular Wireless Networks. In 2011, she joined China Mobile as its Chief Scientist of wireless technologies, established the Green Communications Research Center, and launched the 5G Key Technologies R&D. She is spearheading major initiatives including 5G, C-RAN, high energy efficiency system architectures, technologies and devices; and green energy. She was an Area Editor of *IEEE/ACM Trans. NET*, an elected Board Member of IEEE ComSoc, Chair of the ComSoc Meetings and Conferences Board, and Founding Chair of the IEEE WCNC Steering Committee. She was a Professor at NCTU, an Adjunct Professor at NTU, and currently an Adjunct Professor at BUPT. She is the Chair of FuTURE 5G SIG, an Executive Board Member of GreenTouch, a Network Operator Council Founding Member of ETSI NFV, a Steering Board Member of WWRF, a member of IEEE ComSoc SDB, SPC, and CSCN-SC, and a Scientific Advisory Board Member of Singapore NRF. Her current research interests center around Green, Soft, and Open.

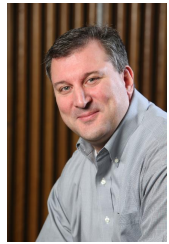

**Thierry E. Klein** is currently the Head of the Innovation Management Program for Vertical Industries within the Nokia Innovation Steering organization at Nokia. Prior to his current role, Thierry was the Program Leader for the Network Energy Research Program at Bell Labs, Alcatel-Lucent with the mission to conduct research towards the design, development and use of sustainable future communications and data networks. He also served as the Chairman of the Technical Committee of GreenTouch, a global consortium dedicated to improve energy efficiency in networks by a factor 1000x compared to 2010 levels. Since 2014, he is also a member of the Momentum for Change Advisory Panel of the UN Framework Convention for Climate Change (UNFCCC). He currently also serves as the Co-Chair of the IEEE Green ICT Initiative.

He joined Bell Labs Research in Murray Hill, New Jersey as a Member of Technical Staff in 2001 conducting fundamental and applied research on next-generation wireless and wireline networks, network architectures, algorithms and protocols, network management, optimization and control. From 2006 to 2010 he served as the Founder and CTO of an internal start-up focused on wireless communications for emergency response and disaster recovery situations within Alcatel-Lucent Ventures.

He earned an MS in Mechanical Engineering and an MS in Electrical Engineering from the Universit de Nantes and the Ecole Centrale de Nantes in Nantes, France. Thierry received a PhD in Electrical Engineering and Computer Science from the Massachusetts Institute of Technology, USA. He is an author on over 35 peer-reviewed conference and journal publications and an inventor on 36 patent applications. He is the recipient of a Bell Labss President Award and two Bell Labs Teamwork Awards. In 2010, he was voted Technologist of the Year at the Total Telecom World Vendor Awards.

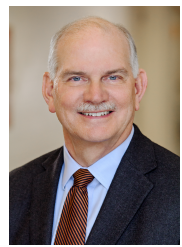

**H. Vincent Poor** (S72, M77, SM82, F87) received the Ph.D. degree in EECS from Princeton University in 1977. From 1977 until 1990, he was on the faculty of the University of Illinois at Urbana-Champaign. Since 1990 he has been on the faculty at Princeton, where he is the Michael Henry Strater University Professor of Electrical Engineering and Dean of the School of Engineering and Applied Science. His research interests are in the areas of information theory, statistical signal processing and stochastic analysis, and their applications in wireless networks and related fields. Among his publications in these areas are the recent books *Principles of Cognitive Radio* (Cambridge University Press, 2013) and *Mechanisms and Games for Dynamic Spectrum Allocation* (Cambridge University Press, 2014).

Dr. Poor is a member of the National Academy of Engineering, the National Academy of Sciences, and is a foreign member of Academia Europaea and the Royal Society. He is also a fellow of the American Academy of Arts and Sciences and other national and international academies. He received the Marconi and Armstrong Awards of the IEEE Communications Society in 2007 and 2009, respectively. Recent recognition of his work includes the 2014 URSI Booker Gold Medal, the 2015 EURASIP Athanasios Papoulis Award, the 2016 John Fritz Medal, and honorary doctorates from Aalborg University, Aalto University, the Hong Kong University of Science and Technology, and the University of Edinburgh.

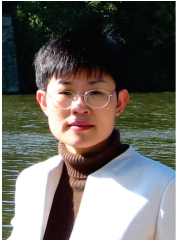

**Chenyang Yang** received her Ph.D. degree in Electrical Engineering from Beihang University (formerly Beijing University of Aeronautics and Astronautics), Beijing, China, in 1997. She has been a full professor with the School of Electronics and Information Engineering, Beihang University since 1999. Dr. Yang was nominated as an Outstanding Young Professor of Beijing in 1995 and was supported by the 1st Teaching and Research Award Program for Outstanding Young Teachers of Higher Education Institutions by Ministry of Education of China during 1999-2004. Dr. Yang was the chair of the IEEE Communications Society Beijing chapter during 2008-2012. She has served as Technical Program Committee Member for numerous IEEE conferences. She has been an associate editor or guest editor of several IEEE journals. Her recent research interests include green radio, local caching, and other emerging techniques for next generation wireless networks.

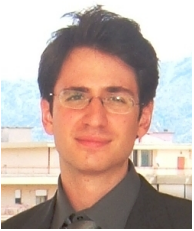

**Alessio Zappone** (S'08 - M'11) is a research associate at the Technische Universität Dresden, Dresden, Germany. Alessio received his M.Sc. and Ph.D. both from the University of Cassino and Southern Lazio. Afterwards, he worked with Consorzio Nazionale Interuniversitario per le Telecomunicazioni (CNIT) in the framework of the FP7 EU-funded project TREND, which focused on energy efficiency in communication networks. Since 2012, Alessio is the project leader of the project CEMRIN on energy-efficient resource allocation in wireless networks, funded by the German research foundation (DFG).

His research interests lie in the area of communication theory and signal processing, with main focus on optimization techniques for resource allocation and energy efficiency maximization. He held several research appointments at TU Dresden, Politecnico di Torino, Supelec - Alcatel-Lucent Chair on Flexible Radio, and University of Naples Federico II. He was the recipient of a Newcom# mobility grant in 2014. Alessio currently serves as associate editor for the IEEE SIGNAL PROCESSING LETTERS.